\documentclass[preprint,onecolumn,nofootinbib]{revtex4}
\pdfoutput=1

\usepackage[colorlinks=true,linkcolor=blue,urlcolor=blue,filecolor=black,citecolor=red,pdfstartview=FitV,pdftitle={},pdfsubject={},pdfkeywords={},pdfpagemode=None,bookmarksopen=true]{hyperref}
\usepackage{graphicx}
\usepackage{amsmath}
\usepackage{amsfonts}
\usepackage{amssymb,ulem}
\usepackage{color}%
\usepackage{dcolumn}
\newcommand{\f}{\begin{equation}}
\newcommand{\ff}{\end{equation}}
\newcommand{\fa}{\begin{eqnarray}}
\newcommand{\ffa}{\end{eqnarray}}

\begin{document}
\title{Holographic transports from Born-Infeld electrodynamics with momentum dissipation}
\author{Xiao-Mei Kuang$^{1}$}
\email{xmeikuang@gmail.com}
\author{Jian-Pin Wu$^{1,3}$}
\email{jianpinwu@yzu.edu.cn}
\author{Zhenhua Zhou$^{2}$}
\email{dtplanck@163.com}
\affiliation{\small{$^1$~Center for Gravitation and Cosmology, College of Physical Science and Technology, Yangzhou University, Yangzhou 225009, China}}
\affiliation{\small{$^2$~School of Physics and Electronic Information, Yunnan Normal University, Kunming, 650500, China}}
\affiliation{\small{$^3$~Institute of Gravitation and Cosmology, Department of
Physics, School of Mathematics and Physics, Bohai University, Jinzhou 121013, China}}

\begin{abstract}
We construct the Einstein-axions AdS black hole from Born-Infeld electrodynamics.
Various DC transport coefficients of the dual boundary theory are computed. The DC electric conductivity depends on the temperature, which is a novel property comparing to that in RN-AdS black hole. The DC electric conductivity are positive at zero temperature while the thermal conductivity vanishes, which implies that the dual system is an electrical metal but thermal insulator. The effects of Born-Infeld parameter on the transport coefficients are analyzed. Finally, we study the AC electric conductivity from Born-Infeld electrodynamics with momentum dissipation. For weak momentum dissipation, the low frequency behavior satisfies the standard Drude formula and the electric transport is coherent
for various correction parameter. While for stronger momentum dissipation,  the
modified Drude formula is applied and we observe a crossover from coherent to incoherent
phase. Moreover, the Born-Infeld correction amplifies the incoherent behavior. Finally, we study the non-linear conductivity in probe limit and compare our results with those observed in (i)DBI model.
\end{abstract}
\maketitle
\section{Introduction}
The gauge-gravity duality \cite{Maldacena,Gubser,Witten} provides a new avenue to study strongly coupled systems, which is difficult to process in the traditional perturbation theory. As an implement of this holographic application, transport phenomenon attracts lots of concentration by studying the electric-thermo linear response via gauge-gravity duality.
In the study, the introduction of momentum relaxation is required to describe more real condensed matter systems, such that finite DC transport coefficients can be realized. In order to include momentum relaxation in the dual theory, several ways are proposed in the
bulk gravitational sectors.

A simple mechanism to introduce the momentum dissipation is in the massive gravity framework.
In this mechanism, the momentum dissipation in the dual boundary field theory is implemented by breaking the diffeomorphism invariance in the bulk \cite{Vegh:2013sk}.
It inspired remarkable progress in holographic studies with momentum relaxation in massive gravity \cite{Davison:2013jba,Blake:2013owa,Blake:2013bqa,Amoretti:2014mma,Zhou:2015dha,Mozaffara:2016iwm,Baggioli:2014roa,Alberte:2015isw,
Baggioli:2016oqk,Zeng:2014uoa,Amoretti:2014zha,Amoretti:2015gna,Fang:2016wqe,Kuang:2017rpx}.

Another mechanism is to introduce a spatial-dependent source, which breaks the Ward identity and the momentum is not conserved in the dual  boundary theory.
An obvious way is the so called scalar lattice or ionic lattice structure, which is implemented by a periodic scalar source or chemical potential \cite{Horowitz:2012ky,Horowitz:2012gs,Ling:2013nxa}.
Also, we can obtain the boundary spontaneous modulation profiles in some particular gravitational models, which break the translational symmetry and induce the charge, spin or pair density waves \cite{Aperis:2010cd,Donos:2013gda,Ling:2014saa,Cremonini:2016rbd,Cai:2017qdz}.
These ways involve solving partial differential equations (PDEs) in the bulk.
Another important way is to break the translation symmetry but hold the homogeneity of the background geometry.
Comparing with the scalar lattice or ionic lattice structure,
the advantage of this way is that we only need to solve ordinary differential equations (ODEs) in the bulk.
Outstanding examples of this include holographic Q-lattices \cite{Donos:2013eha,Donos:2014uba,Ling:2015epa,Ling:2015exa}, helical lattices \cite{Donos:2012js} and axion model \cite{Andrade:2013gsa,Kim:2014bza,Cheng:2014tya,Ge:2014aza,Andrade:2016tbr,Kuang:2017cgt,Cisterna:2018hzf,Kuang:2016edj,Tanhayi:2016uui,Cisterna:2017jmv,Cisterna:2017qrb}.
Holographic Q-lattice model breaks the translational invariance via the global phase of the complex scalar field.
Holographic helical lattice model possesses the non-Abelian Bianchi VII$_0$ symmetry, where the translational symmetry is broken in one space direction but holds in the other two space directions.
The translational symmetry is broken in holographic axion model by
a pair of spatial-dependent scalar fields, which are introduced to source the breaking of Ward identity.
In addition, by turning on a higher-derivative interaction term between the $U(1)$ gauge field and the scalar field, we can also obtain a spatially dependent
profile of the scalar field generated spontaneously, which leads to the breaking of the Ward identity and the momentum dissipation in the dual boundary field theory \cite{Kuang:2013oqa,Alsup:2013kda}.

On the other hand,  many non-linear electrodynamics (NLED) has been proposed in the bulk theory instead of the Standard Maxwell (SM) theory due to the two aspects. One is that the SM theory face some problems, for instances, infinite self-energy of point-like charges
, vacuum polarization in quantum electrodynamics and low-energy limit of heterotic string theory \cite{Kats:2006xp,Anninos,Seiberg:1999vs}. The other is as pointed out in \cite{Hassaine:2007py} that in higher dimensions, the action for Maxwell field does not have the conformal symmetry. Among the NLED theory, the pioneering non-linear generalization of the Maxwell theory was proposed in \cite{Born:1934} with the form
\begin{eqnarray}
\label{LM}
\mathcal{L}_{BI}=\frac{1}{\gamma}\left(1-\sqrt{1+\frac{\gamma}{2}F^2}\right)\,,
\end{eqnarray}
which is natural in string theory \cite{Gibbons:2001gy}.
Related holographic study with the Born-Infeld (BI) correction on the Maxwell field can been seen in \cite{Jing:2010zp,Sheykhi:2015mxb,Ghorai:2015wft,Lai:2015rva,Chaturvedi:2015hra,Bai:2012cx,
Gangopadhyay:2012np,Gangopadhyay:2012am,Jing:2010cx,Wu:2016hry,Guo:2017bru,Charmousis:2010zz} and therein, in which the correction introduces interesting properties\footnote{ More forms for non-linear Maxwell theory, such as power Maxwell theory,  logarithmic Maxwell theory were also proposed in \cite{Hassaine:2007py,Soleng:1995kn,Mu:2017usw}.
Also, the magnetotransport in holographic DBI (BI) model has been studied in \cite{Cremonini:2017qwq}.
In particular, in \cite{Kiritsis:2016cpm}, they find that the in-plane magneto-resistivity exhibits the interesting scaling behavior that is compatible
to that observed recently in experiments on $BaFe_2(As_{1-x}P_x)_2$ \cite{analytis}.}.

In this paper, we will study the Einstein-axions theory with Born-Infeld Maxwell field, i.e., the Einstein-Born-Infeld-axions theory.
We first construct the black brane solution by solving the equations of motion in the theory.
Then we analytically compute the DC transport coefficients in the dual theory and we discuss the influence from Born-Infeld parameter.
Also, we numerically study the AC electric conductivity and analyze its low frequency behavior via (modified) Drude formula.
Finally, we analyze the non-linear current-voltage behavior with BI correction in probe limit.

\section{Einstein-Born-Infeld-axions Theory}\label{EBIa}

Since the Born-Infeld Anti de-Sitter (BI-AdS) geometry and its extensions have been explored
in detail in \cite{Dey:2004yt,Cai:2004eh,Cai:2008in,Banerjee:2011cz,Liu:2011cu,Chaturvedi:2015hra} and references therein,
here, we only give a brief review on the BI-AdS geometry related with our present study.

The action in Einstein-Born-Infeld-axions theory we consider is
\begin{eqnarray}
\label{action}
S=\frac{1}{2\kappa^2}\int d^{4}x \sqrt{-g}
\left(R+\frac{6}{L^2}-\frac{1}{2}\sum_{I=x,y}(\partial \phi_I)^2
+\mathcal{L}_{BI}
\right)\,,
\end{eqnarray}
where $\mathcal{L}_{BI}$ is defined in \eqref{LM} and $\phi_I$ is the massless axion fields.
When $\gamma\rightarrow 0$, we have $\mathcal{L}_{BI}=-\frac{1}{4}F^2$, which is the action for the standard  Maxwell theory,
whereas in the limit of $\gamma\rightarrow\infty$, $\mathcal{L}_{BI}=0$,
then our theory (\ref{action}) reduces to Einstein-axions one.

The equations of motion can be straightforward derived from the action (\ref{action}) as follows
\begin{eqnarray}
&&
\Box\phi_I=0
\,,
\label{phiE}
\\
&&
\nabla_{\mu}\Big(\frac{F^{\mu\nu}}{\sqrt{1+\frac{\gamma}{2}F^2}}\Big)=0
\,,
\label{ME}
\\
&&
R_{\mu\nu}-\frac{1}{2}R g_{\mu\nu}-\frac{3}{L^2}g_{\mu\nu}
+\frac{1}{2}T^{(A)}_{\mu\nu}
+\frac{1}{2}T^{(\phi)}_{\mu\nu}
=0\,,
\label{EE}
\end{eqnarray}
where
\fa
&&
\label{TA}
T^{(A)}_{\mu\nu}=-g_{\mu\nu}\mathcal{L}_{BI}-\frac{1}{\sqrt{1+\frac{\gamma}{2}F^2}}F_{\mu\rho}F_{\nu}^{\ \rho}
\
\\
&&
\label{Tphi}
T^{(\phi)}_{\mu\nu}=-\partial_\mu\phi_x\partial_\nu\phi_x
-\partial_\mu\phi_y\partial_\nu\phi_y
+\frac{g_{\mu\nu}}{2}(\partial \phi_x)^2
+\frac{g_{\mu\nu}}{2}(\partial \phi_y)^2\,.
\ffa

The model (\ref{action}) supports an AdS$_4$ solution with AdS radius $L$,
which shall be set $L=1$ in what follows.
We are interested in the homogeneous and isotropic charged black brane solution
with spatial linear dependent scalar field. Then we set the fields as
\fa
\label{metric}
ds^2=\frac{1}{u^2}\Big(-f(u)dt^2+\frac{du^2}{f(u)}+\delta_{ij}dx^idx^j\Big)\,,~~~~~~~~
A=A_t(u)dt\,,~~~~~~~~\phi_I=\beta_{Ii}x^i\,,
\ffa
where $i$ denotes the $2$ spatial $x^i$ directions,
$I$ is an internal index that labels the $2$ scalar fields and
$\alpha_{Ii}$ are real arbitrary constants.
Notice that in the above ansatz, we work in the coordinate system
in which $u\rightarrow 0$ is the UV boundary and $u=1$ denotes the location of horizon.
The equations of motion (\ref{phiE}), (\ref{ME}) and (\ref{EE}) give
\fa
&&
\label{fr}
f(u)=
\frac{\mu ^2 u^4 \, _2F_1\left(\frac{1}{4},\frac{1}{2};\frac{5}{4};-u^4 \gamma  \mu ^2\right)}{3}
+\frac{1-\sqrt{\gamma  \mu ^2 u^4+1}}{6 \gamma}-M u^3-\frac{\beta ^2 u^2}{2}+1\,,
\
\\
&&
\label{Atr}
A_t(u)=\mu_{BI}\left(1
-u \frac{\, _2F_1\left(\frac{1}{4},\frac{1}{2};\frac{5}{4};-\gamma \mu ^2u^4\right)}{_2F_1\left(\frac{1}{4},\frac{1}{2};\frac{5}{4};-\gamma\mu^2\right)}\right)\,,
\ffa
where
\fa
\mu_{BI}=
\mu~_2F_1\left(\frac{1}{4},\frac{1}{2};\frac{5}{4};-\gamma\mu^2\right)\,
\ffa
is the chemical potential of the system and $\mu_{BI}=\mu$ as $\gamma\rightarrow 0$.
$M$ is determined by $f(u=1)=0$ at the horizon as
\fa
M=
-\frac{\beta ^2}{2}-\frac{\sqrt{\gamma  \mu ^2+1}}{6 \gamma}+\frac{1}{6 \gamma}
+\frac{\mu ^2 \, _2F_1\left(\frac{1}{4},\frac{1}{2};\frac{5}{4};-\gamma  \mu ^2\right)}{3}+1
   \,.
\ffa

The Hawking temperature of the black brane is
\fa
\label{tem}
T=\frac{1}{4\pi}\left(3-\frac{\beta ^2}{2}+\frac{1}{2\gamma}\left(1-\sqrt{\gamma  \mu ^2+1}\right)\right)\,.
\ffa
This black brane solution is specified by the two dimensionless parameters $\hat{T}=\frac{T}{\mu_{BI}}$ and $\hat{\beta}=\frac{\beta}{\mu_{BI}}$,
in which the temperature can be reexpressed as
\fa
\label{temr}
\hat{T}=\frac{-\hat{\beta} ^2 \gamma  \mu _{\text{BI}}^2-\sqrt{\gamma  \mu ^2+1}+6 \gamma+1}{8 \pi
   \gamma  \mu _{\text{BI}}}\,.
\ffa

Now, we have obtained an analytical black brane solution in the framework of Einstein-Born-Infeld-axions theory.
Notice that when $\gamma\to 0$, the non-linear action for Maxwell field \eqref{LM} can be expanded into the hand-given form equation (2.9) with tiny $\Theta$ in \cite{Baggioli:2016oju}.
And they discuss the case $\Theta>0$ (corresponding to $\gamma<0$ here) to address the insulating phase.
But when $\gamma<0$, there is a value of $\gamma$, below which the black brane solution becomes complex.
In this paper, we shall mainly focus on the holographic properties of all DC transport coefficients and AC electric conductivity at low frequency region.
So we only consider $\gamma>0$ unless we specially point out.

\section{Electric, thermal and thermoelectric DC conductivity}

In this section, we will calculate the DC conductivity
including electric, thermal and thermoelectric conductivity via the technics proposed in \cite{Donos:2014uba,Donos:2014cya,Blake:2014yla}.
To this end, we consider the following consistent perturbations at the linear level
\begin{align}
\label{DCperturbations}
&\delta g_{tx}=\frac{1}{u^2}(H(u)t+h_{tx}(u))\,,~~~\delta g_{ux}=\frac{1}{u^2}h_{ux}(u)\,,\nonumber\\
&\delta A_{x}=E_p(u)t+a_x(u)\,,~~~~~~~~~\delta \phi_{x}=\delta \chi_{x}(u)\,.
\end{align}
According to \cite{Donos:2014cya}, one defines two radial conserved quantities whose values at the boundary ($u\rightarrow 0$) are related respectively to the charge and heat response currents in the dual field,
\begin{align}
\mathcal{J}=\sqrt{-g}\frac{F^{ux}}{\sqrt{1+\frac{\gamma}{2}F^2}}\,,~~~~~~
\mathcal{J}^Q=2\sqrt{-g}\nabla^uk^x-A_t\mathcal{J}\,,
\end{align}
where $k^\mu=\partial^t$ is the Killing vector. In terms of the background ansatz \eqref{metric} and the perturbation \eqref{DCperturbations}, the two conserved currents read explicitly as
\begin{align}
&\mathcal{J}=\frac{t H A'_t+h_{tx} A'_t+f \left(a'_x+tE'_p\right)}{\sqrt{1-\gamma  u^4 (A'_t)^2}}\,,\\
&\mathcal{J}^Q=-\frac{-t H f'-h_{tx} f'+f (t H'+h'_{tx})}{u^2}-A_t\mathcal{J}\,.
\end{align}
We assume the special forms of $E_p(u)=-E_x+\zeta A_t(u),\,H(u)=-\zeta f(u)$ where the constants $E_x$
and $\zeta$ parametrize the sources for the electric current and heat current, respectively. Then, the related terms with respect to the time $t$ can be canceled and the conserved currents become
\begin{align}
\mathcal{J}=-Q_{BI}\Big(h_{tx}+\frac{f a'_x}{A'_t}\Big)\,,~~~~~~
\mathcal{J}^Q=-\frac{f^2}{u^2}\Big(\frac{h_{tx}}{f}\Big)'-A_t\mathcal{J}\,.
\end{align}
In the above expression, we have defined the charge density $Q_{\text{BI}}$ as
\fa
Q_{\text{BI}}=-\frac{A_t'(u)}{\sqrt{1-\gamma  u^4 A_t'(u)^2}}\,,
\ffa
which is the conserved electric charge density.

Next, we shall evaluate the DC conductivities by the following expressions\cite{Donos:2014cya}
\begin{align}
\label{DCexp}
\sigma_{DC}=\frac{\partial \mathcal{J}}{\partial E_x}\,,~~~~
\bar{\alpha}_{DC}=\frac{1}{T}\frac{\partial \mathcal{J}^Q}{\partial E_x}\,,~~~
\alpha_{DC}=\frac{1}{T}\frac{\partial \mathcal{J}}{\partial \xi}\,,~~~
\bar{\kappa}_{DC}=\frac{1}{T}\frac{\partial \mathcal{J}^Q}{\partial \xi}\,.
\end{align}
Since $\mathcal{J}$ and $\mathcal{J}^Q$ are both conserved quantities along $u$ direction,
we can evaluate the above expressions at horizon.
To achieve this goal, we analyze the behaviors of the perturbative quantities at the horizon.
First, it is easy to obtain the following express from Einstein equation
\begin{align}
\label{hux}
h_{ux}=\frac{-Q_{BI} E_p +\beta f \chi'_x+H'}{\beta^2 f}\,.
\end{align}
Further, we have
\begin{align}
\label{htx-ax}
h_{tx}= -fh_{ux}=-\frac{Q_{BI} E_x }{\beta^2 }+\zeta\frac{f'}{\beta^2}\,,~~~~f^2\Big(\frac{h_{tx}}{f}\Big)'=\frac{Q_{BI} E_xf' }{\beta^2 }-\zeta\frac{(f')^2}{\beta^2}\,,~~~~a'_x=\frac{E_x}{f}\,.
\end{align}
Note that the above equations including Eqs.\eqref{hux} and \eqref{htx-ax} have taken value at the horizon, i.e., $u=1$.
And then we can evaluate the currents at the horizon, which give
\begin{align}
\mathcal{J}=E_x\Big(-\frac{Q_{BI}}{A'_t}+\frac{Q_{BI}^2}{\beta^2}\Big)-\zeta\frac{Q_{BI}f'}{\beta^2}\,,~~~~~~
\mathcal{J}^Q=-E_x\frac{Q_{BI}f'}{\beta^2 }+\zeta\frac{(f')^2}{\beta^2}\,.
\end{align}

Thus the conductivities computed from \eqref{DCexp} can be expressed as
\begin{align}
&\sigma_{DC}=\frac{\partial \mathcal{J}}{\partial E_x}=\sqrt{1+\gamma \mu^2}+\frac{\mu^2}{\hat{\beta}^2\mu_{BI}^2}
\label{sigmaDC}
\,,\\
&\bar{\alpha}_{DC}=\frac{1}{T}\frac{\partial \mathcal{J}^Q}{\partial E_x}=\frac{4 \pi \mu}{\hat{\beta}^2\mu_{BI}^2}
\label{alphabDC}
\,,\\
&\alpha_{DC}=\frac{1}{T}\frac{\partial \mathcal{J}}{\partial \xi}=\frac{4 \pi \mu}{\hat{\beta}^2\mu_{BI}^2}
\label{alphaDC}
\,,\\
&\bar{\kappa}_{DC}=\frac{1}{T}\frac{\partial \mathcal{J}^Q}{\partial \xi}=\frac{(4\pi)^2\hat{T}}{\hat{\beta}^2\mu_{BI}}
\label{kappabDC}
\,.~~~~
\end{align}
Also, we are interested in the thermal conductivity at zero electric current, which is defined as
\fa
\label{kappa}
\kappa_{DC}\equiv\bar{\kappa}_{DC}-\frac{\alpha_{DC}\bar{\alpha}_{DC}T}{\sigma_{DC}}\,,
\ffa
which is more readily measurable than $\bar{\kappa}$. Subseqently, it can be explicitly evaluated as
\fa
\label{kappa-eva}
\kappa_{DC}=\frac{16 \pi ^2 \hat{T} \mu _{\text{BI}} \sqrt{\gamma  \mu ^2+1}}{\hat{\beta }^2 \mu _{\text{BI}}^2
   \sqrt{\gamma  \mu ^2+1}+\mu ^2}\,.
\ffa
When $\gamma\rightarrow0$, all the transport coefficients are coincide with those in Einstein-Maxwell-axions theory studied in \cite{Donos:2014cya}.

\begin{figure}
\center{
\includegraphics[scale=0.55]{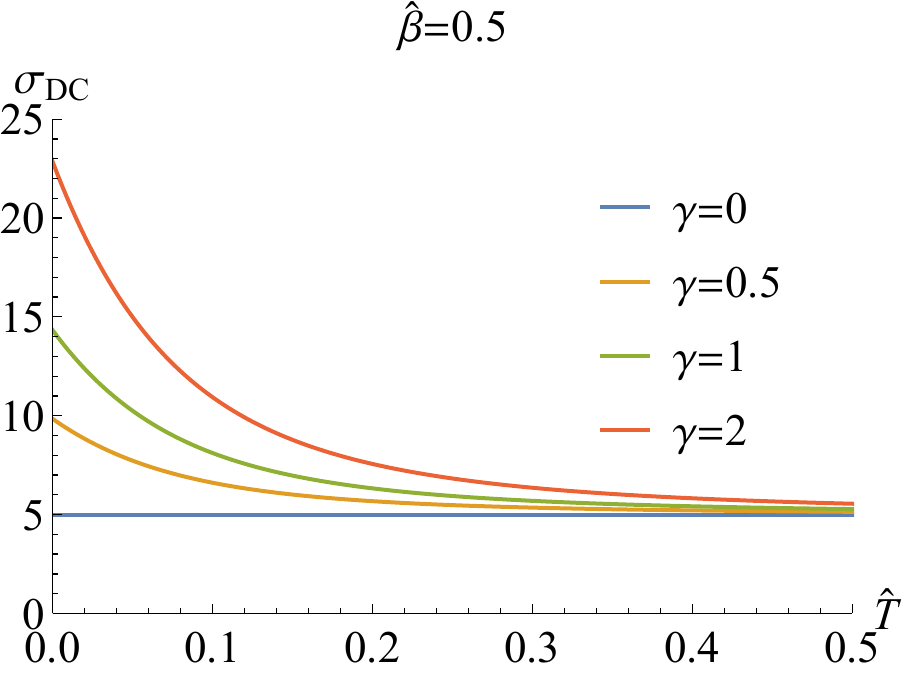}\ \hspace{0.8cm}
\includegraphics[scale=0.55]{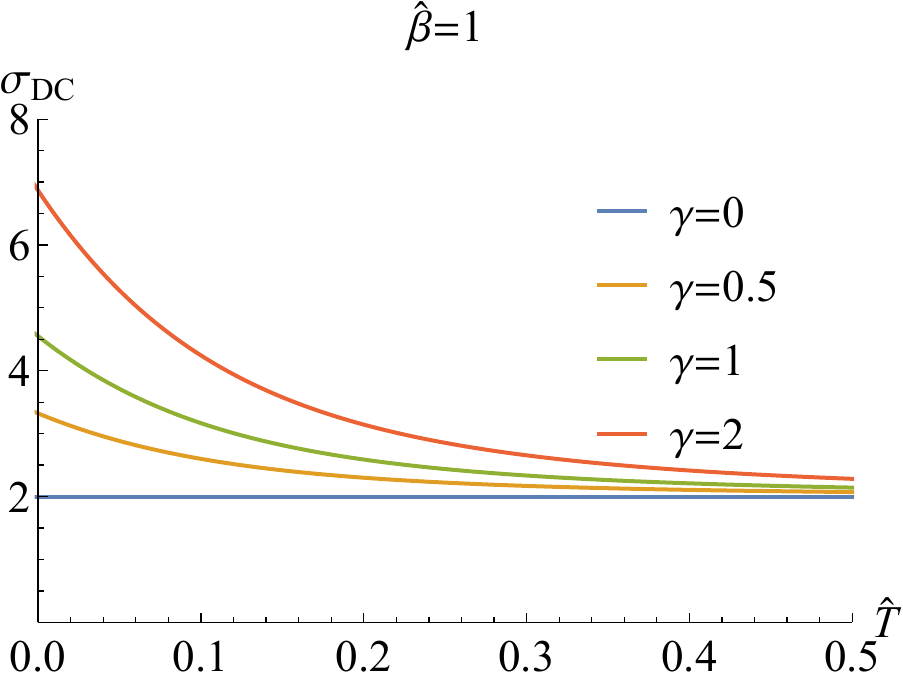}\ \hspace{0.8cm}\ \\
\caption{\label{dcvsTbeta} $\sigma_{DC}$ as the function of $\hat{T}$ for some specific $\gamma$ and $\hat{\beta}$.}}
\end{figure}

\begin{figure}
\center{
\includegraphics[scale=0.55]{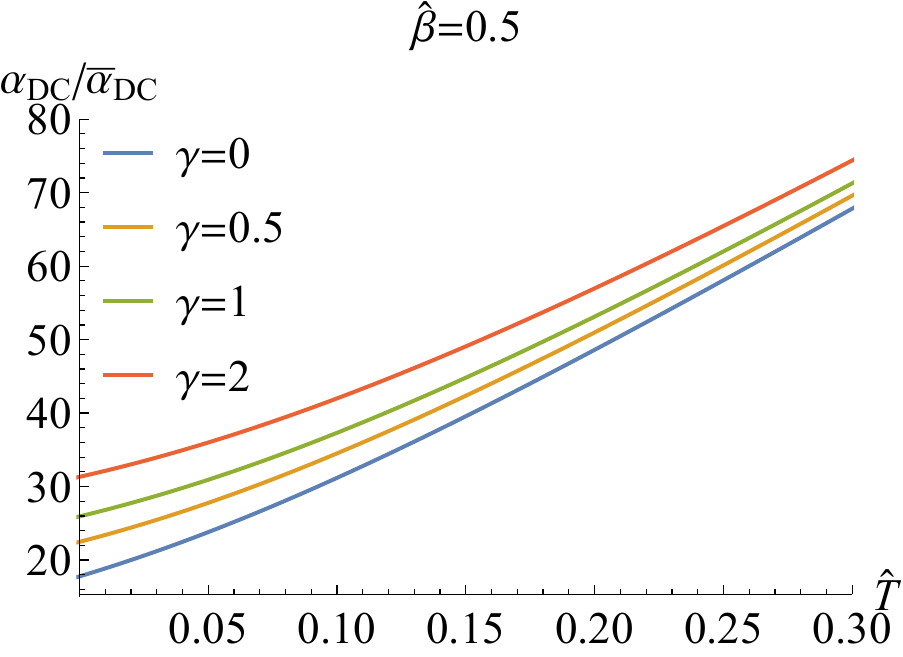}\ \hspace{0.8cm}
\includegraphics[scale=0.55]{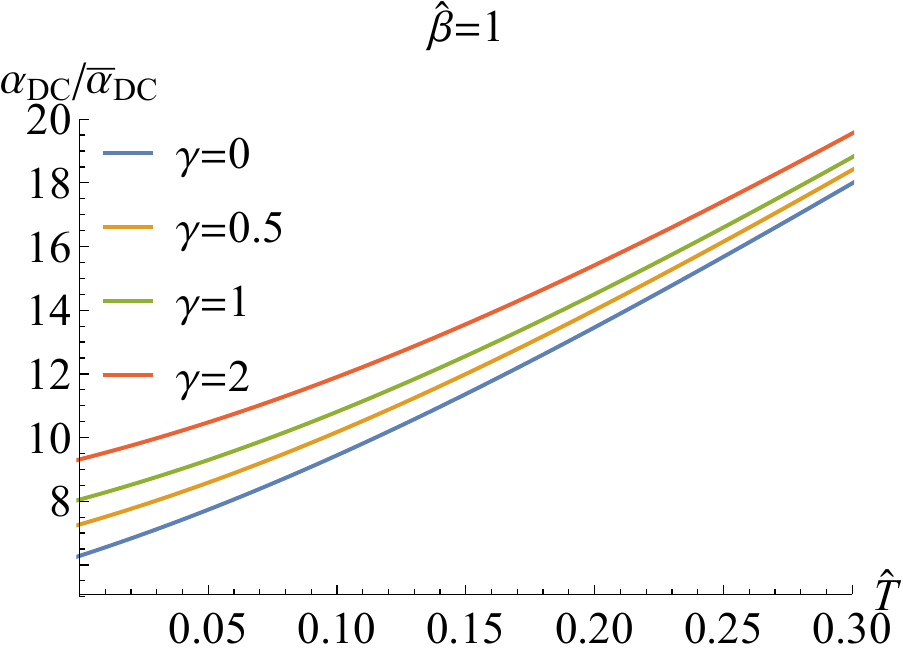}\ \hspace{0.8cm}\ \\
\caption{\label{alpha_dcvsTbeta} $\alpha_{DC}, \bar{\alpha}_{DC}$ as the function of $\hat{T}$ for some specific $\gamma$ and $\hat{\beta}$.}}
\end{figure}

\begin{figure}
\center{
\includegraphics[scale=0.55]{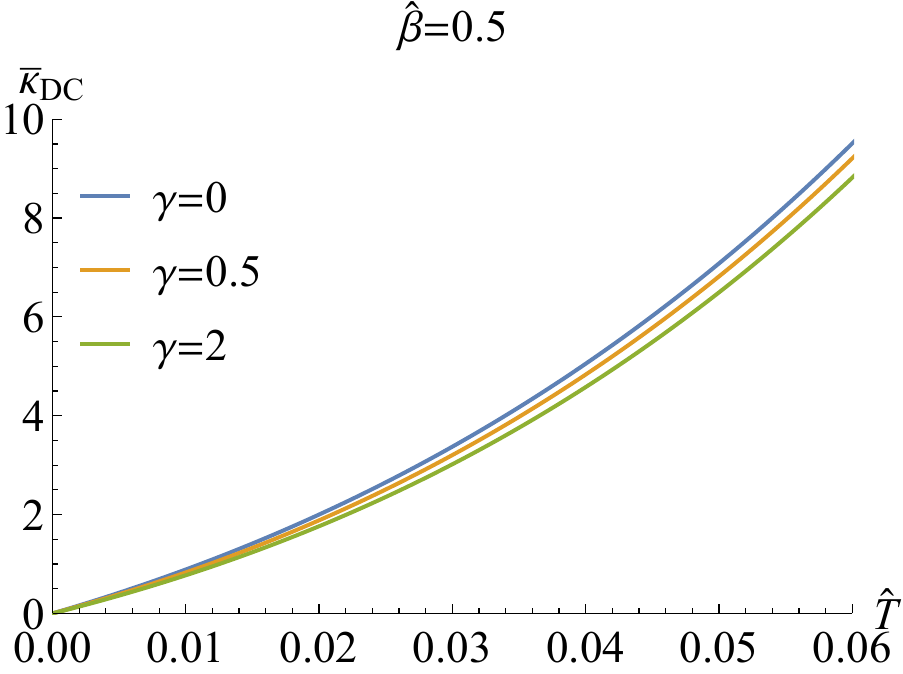}\ \hspace{0.8cm}
\includegraphics[scale=0.55]{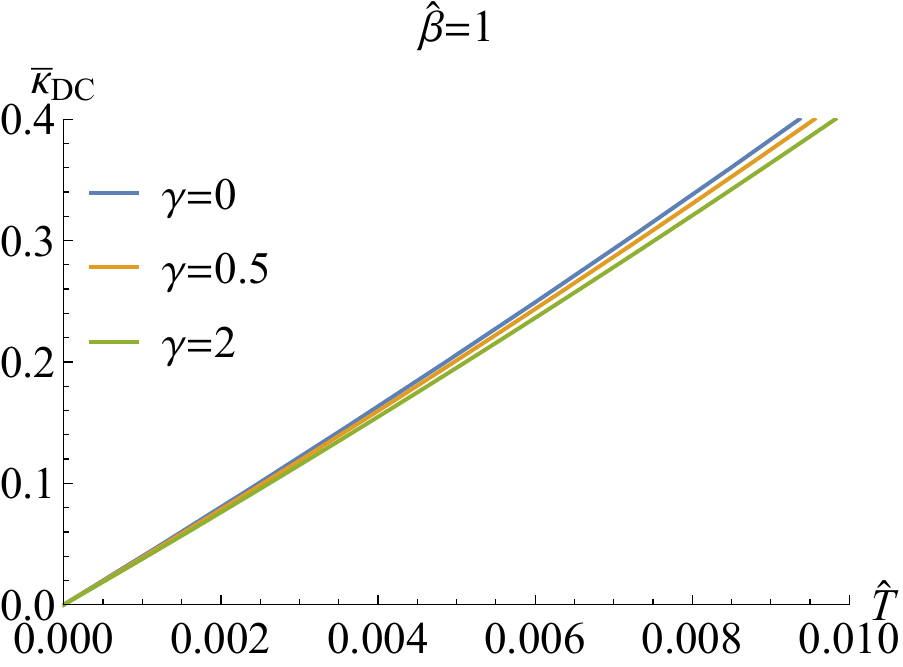}\ \hspace{0.8cm}\ \\
\caption{\label{kappab_dcvsTbeta} $\bar{\kappa}_{DC}$ as the function of $\hat{T}$ for some specific $\gamma$ and $\hat{\beta}$.}}
\end{figure}

\begin{figure}
\center{
\includegraphics[scale=0.55]{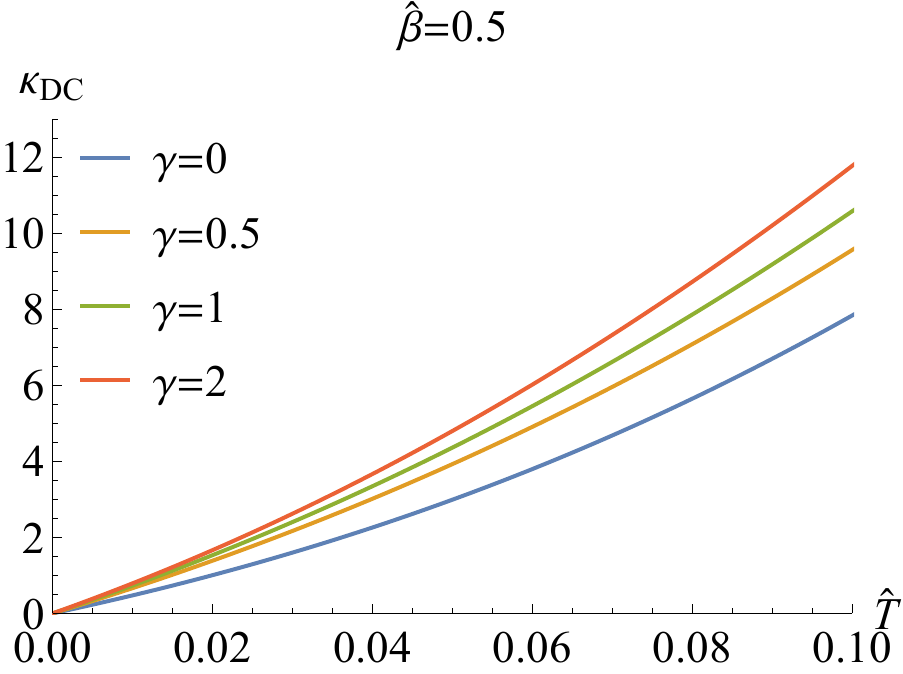}\ \hspace{0.8cm}
\includegraphics[scale=0.55]{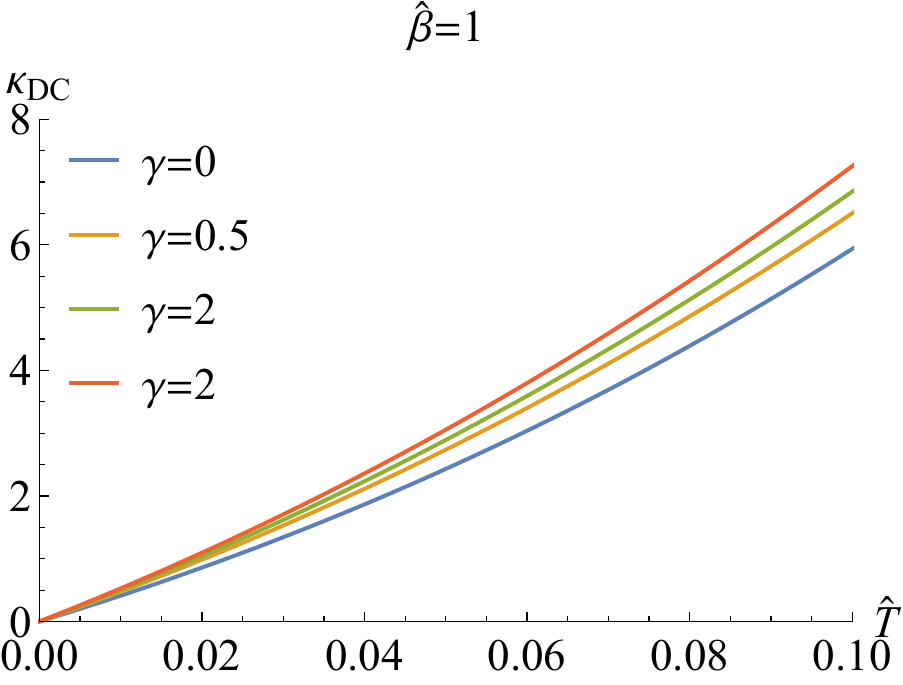}\ \hspace{0.8cm}\ \\
\caption{\label{kappa_dcvsTbeta} $\kappa_{DC}$ as the function of $\hat{T}$ for some specific $\gamma$ and $\hat{\beta}$.}}
\end{figure}

We summarize the main characteristics of the DC conductivities,
\begin{itemize}
  \item The electric DC conductivity $\sigma_{DC}$ is temperature dependent for the fixed $\hat{\beta}$ (FIG.\ref{dcvsTbeta}).
  It is the key difference comparing with that in $4$ dimensional RN-AdS black brane,
in which the DC conductivity is the temperature independence.
\item This system of BI-axions model is an electrical metal  but thermal insulator because at $\hat{T}=0$, $\sigma_{DC}$ is a positive constant (FIG.\ref{dcvsTbeta}) while $\bar{\kappa}=0$ (FIG.\ref{kappab_dcvsTbeta}).
\item For the fixed finite temperature $\hat{T}$, with the increase of $\gamma$, $\sigma_{DC}$,
$\alpha_{DC}/\bar{\alpha}_{DC}$ and $\kappa_{DC}$ increase (FIG.\ref{dcvsTbeta}, \ref{alpha_dcvsTbeta} and \ref{kappa_dcvsTbeta}), but $\bar{\kappa}_{DC}$ decreases (FIG.\ref{kappab_dcvsTbeta}).
\end{itemize}


Another quantities of interest are the Lorentz rations, which are the rations of thermal conductivity to electric conductivity
\fa
&&
\label{Lb}
\bar{L}=\frac{\bar{\kappa}}{\sigma\hat{T}\mu_{\text{BI}}}=\frac{16 \pi ^2}{\hat{\beta }^2 \mu _{\text{BI}}^2 \sqrt{\gamma  \mu ^2+1}+\mu ^2}\,,
\
\\
&&
\label{L}
L=\frac{\kappa}{\sigma\hat{T}\mu_{\text{BI}}}=\frac{16 \pi ^2 \hat{\beta }^2 \mu _{\text{BI}}^2 \sqrt{\gamma  \mu ^2+1}}{\left(\hat{\beta }^2 \mu
   _{\text{BI}}^2 \sqrt{\gamma  \mu ^2+1}+\mu ^2\right){}^2}\,.
\ffa
Obviously, $L$ is not a constant and so the Wiedemann-Franz law that $L$ is a constant for Fermi liquid \cite{Mahajan:2013cja} is violated,
which has been revealed in \cite{Donos:2014cya,Kim:2014bza,Kuang:2017rpx} and indicates that our dual system involves strong interactions.
Similarly with that in holographic Q-lattice model or linear axions model with standard Maxwell theory studied in \cite{Donos:2014cya}, as $\hat{\beta}\rightarrow 0$, $\bar{L}$ and $\kappa$ approach the constants.
It is interesting to notice that in this case, i.e., $\hat{\beta}\rightarrow 0$,
$\bar{L}$ is independent of the BI parameter $\gamma$ but $\kappa$ depends on it,
while $L$ vanishes and $\bar{\kappa}$ diverges in this limit which is similar to that observed in \cite{Donos:2014cya}.
In addition, the bound $\bar{L}\leq\frac{s^2}{Q}$ with $s$ being the entropy density of the black brane proposed in \cite{Donos:2014cya}
holds in our model.

\section{Optical electric conductivity}
In this section, we turn to study the AC electric conductivity by turning on the following consistent frequency dependent perturbations
\fa
\delta A_x=\int\frac{d\omega}{2\pi}e^{-i\omega t}a_x(u)\,,~~\delta g_{tx}=\int\frac{d\omega}{2\pi}e^{-i\omega t}\frac{h_{tx}(u)}{u^2}\,,~~
\delta\phi_x=\int\frac{d\omega}{2\pi}e^{-i\omega t}\chi_x(u)\,.
\ffa
Thus, the linearized equations of motion around the background \eqref{metric} can be derived in momentum space as
\fa
&&
\label{eompax}
a_x'(u) \left(2 \gamma  u^3 A_t'(u){}^2+\frac{f'(u)}{f(u)}\right)+\frac{\omega ^2
   a_x(u)}{f(u)^2}+a_x''(u)+\frac{A_t'(u) h_{\text{tx}}'(u)}{f(u)}
   =0\,,
\
\\
&&
\label{eomphtx}
4 u^2 \omega  a_x(u) A_t'(u)+\sqrt{1-\gamma  u^4 A_t'(u){}^2} \left(\omega  h_{\text{tx}}'(u)-\alpha
   f(u) \chi _x'(u)\right)=0\,,
\
\\
&&
\label{eompchx}
\left(\frac{f'(u)}{f(u)}-\frac{2}{u}\right) \chi _x'(u)-\frac{\alpha  \omega
   h_{\text{tx}}(u)}{f(u)^2}+\frac{\omega ^2 \chi _x(u)}{f(u)^2}+\chi _x''(u)=0\,.
\ffa
According to AdS/CFT dictionary, we can numerically solve the above equations and read off the AC conductivity by using the expression
\fa
\label{ACsigma}
\sigma(\omega)=\frac{\partial_ua_x}{i\omega a_x}\,.
\ffa


\begin{figure}
\center{
\includegraphics[scale=0.6]{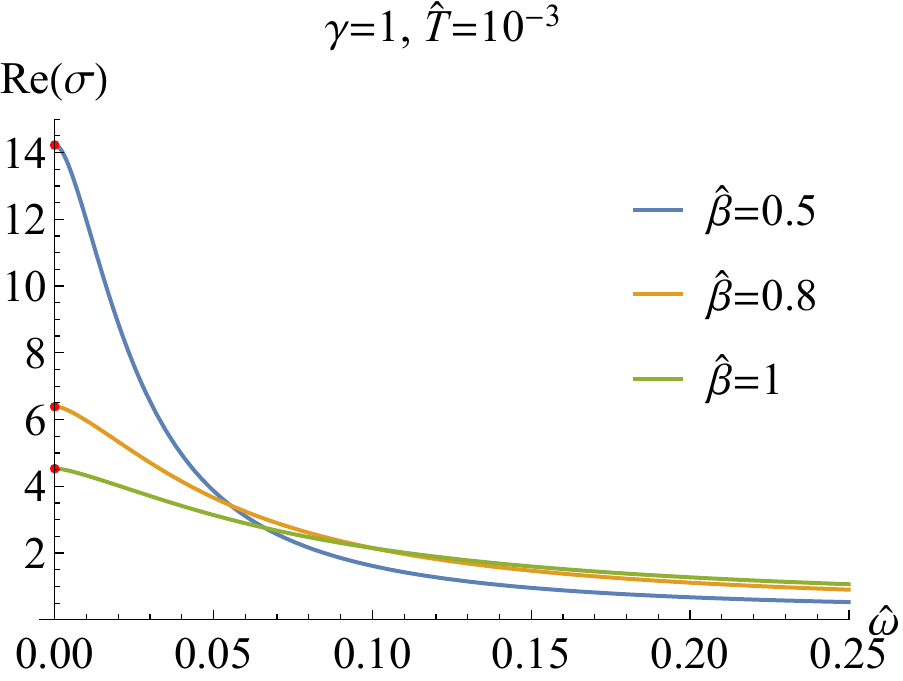}\ \hspace{0.8cm}
\includegraphics[scale=0.6]{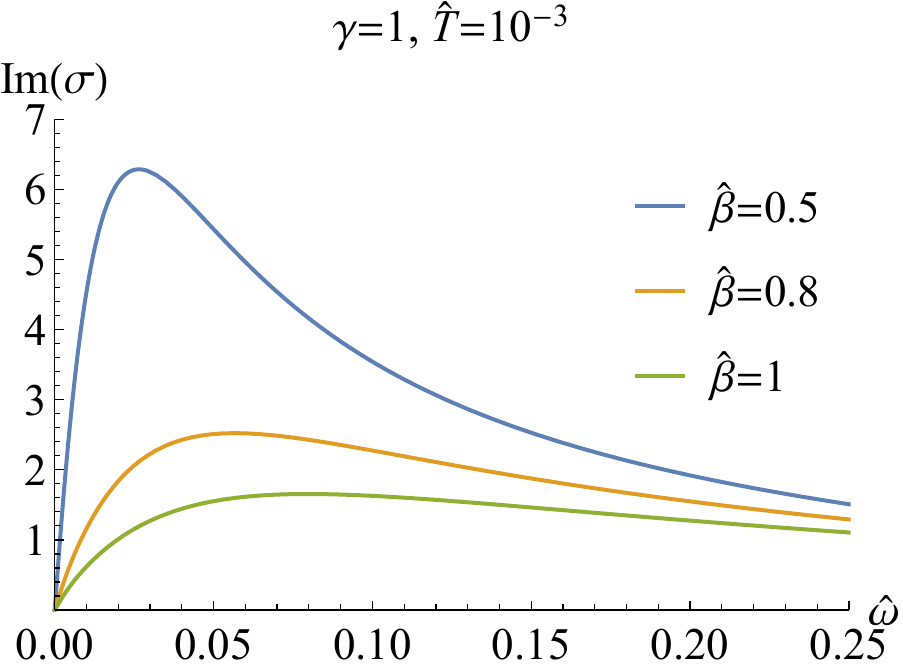}\ \hspace{0.8cm}\ \\
\includegraphics[scale=0.6]{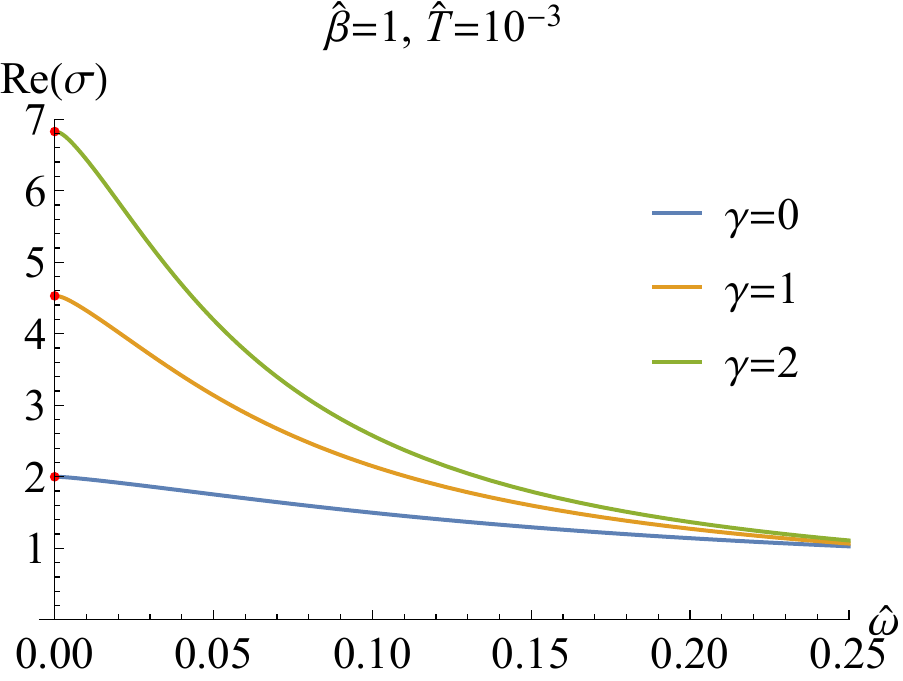}\ \hspace{0.8cm}
\includegraphics[scale=0.6]{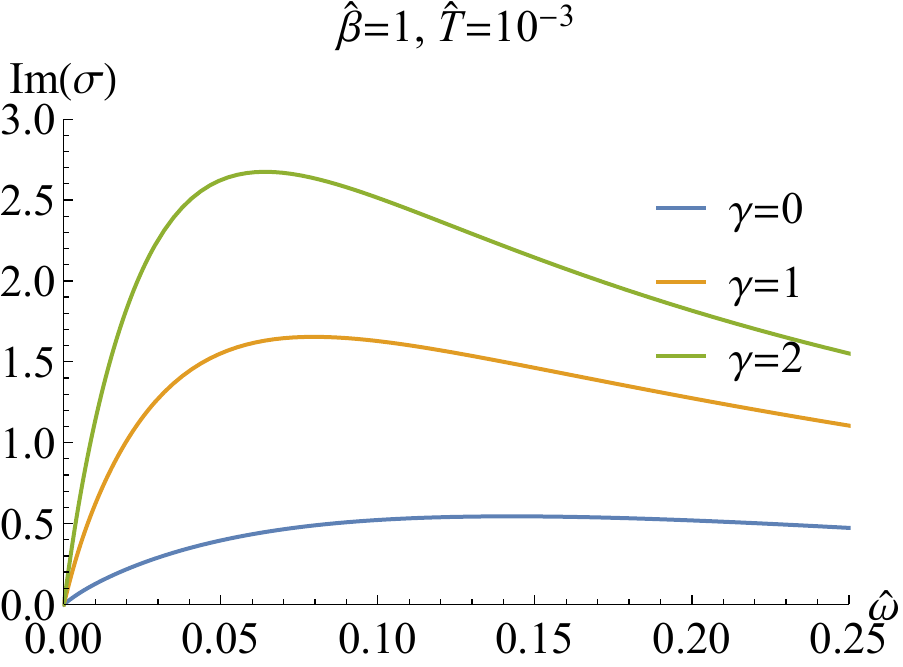}\ \hspace{0.8cm}\ \\
\caption{\label{sigmav1} The electric conductivity $\sigma(\hat{\omega})$ for different $\hat{\beta}$ and $\gamma$.}}
\end{figure}

We will explore the AC electric conductivity from non-linear BI electrodynamics with momentum dissipation.
We shall fix the temperature $\hat{T}$ to study the effects of $\hat{\beta}$ and $\gamma$.
FIG.\ref{sigmav1} exhibits the electric conductivity as the function of the frequency $\hat{\omega}$ for different $\hat{\beta}$ and $\gamma$.
Similarly with the standard Maxwell theory, for the fixed $\gamma$, with the increase of $\hat{\beta}$,
the Drude-like peak gradually reduces and a transition from coherent phase to incoherent phase happens.
For the fixed $\hat{\beta}$, the peak seems to augment when $\gamma$ increases.
But the quantitative analysis later indicates that although with the increase of $\gamma$ the peak augments,
the degree of deviation from the Drude one becomes grave.
Note that as a quick check on the consistency of our numerics, we denote the DC electric conductivity analytically calculated by \eqref{sigmaDC} (red points) in FIG.\ref{sigmav1},
which match very well with the numerical results.

Next we discuss the coherent and incoherent behavior of our present model
by  quantitatively exploring the low frequency behavior of the AC conductivity.
It is well known that for the standard Maxwell theory, when the momentum dissipation is weak,
i.e. $\hat{\beta}\ll 1$, the conductivity at low frequency can be described by the standard Drude formula,
\fa
\label{Drude-s}
\sigma(\hat{\omega})=\frac{K}{\Gamma-i\hat{\omega}}\,,
\ffa
where $K$ is a constant and $\Gamma$ the momentum relaxation rate.
It is the coherent transport.
With the increase of $\hat{\beta}$, there is a crossover from coherent to incoherent phase,
which can be depicted by the following modified Drude formula
\fa
\label{Drude-m}
\sigma(\hat{\omega})=\frac{K}{\Gamma-i\hat{\omega}}+\sigma_Q\,.
\ffa
The above formula can be obtained in relativistic conformal hydrodynamics \cite{Hartnoll:2007ih} and
$\sigma_Q$ is the intrinsic conductivity of the hydrodynamic state, characterizing the incoherent contribution.
In holographic framework, this formula has also been widely applied to study the coherent and incoherent behavior, for example, see \cite{Kim:2014bza,Ling:2015exa,Kuang:2017cgt}.

Here, we shall study the low frequency conductivity behavior by using these two formulas
and intend to give some quantitative discussions and insights into it.
\begin{figure}
\center{
\includegraphics[scale=0.6]{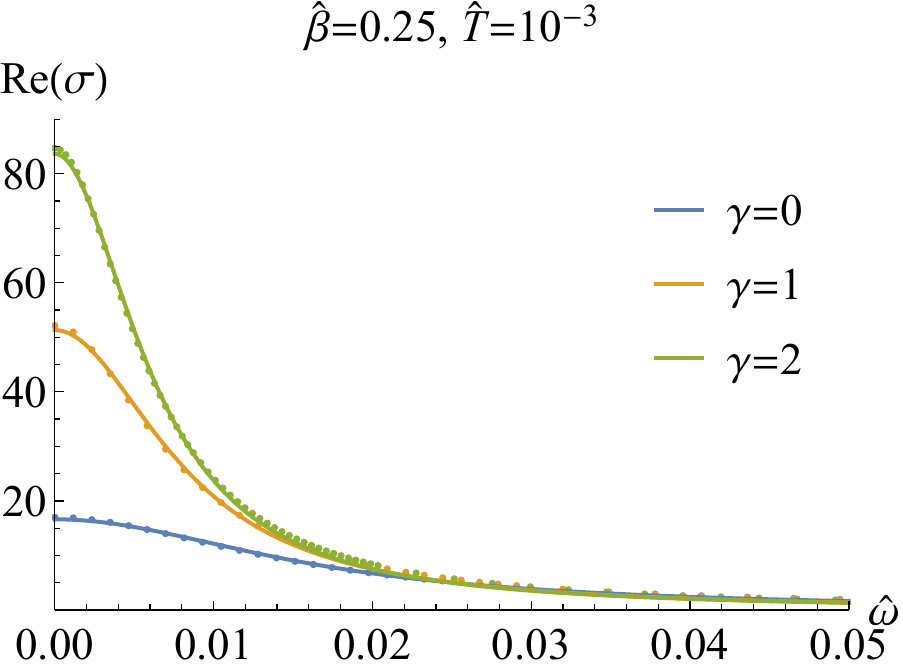}\ \hspace{0.8cm}
\includegraphics[scale=0.6]{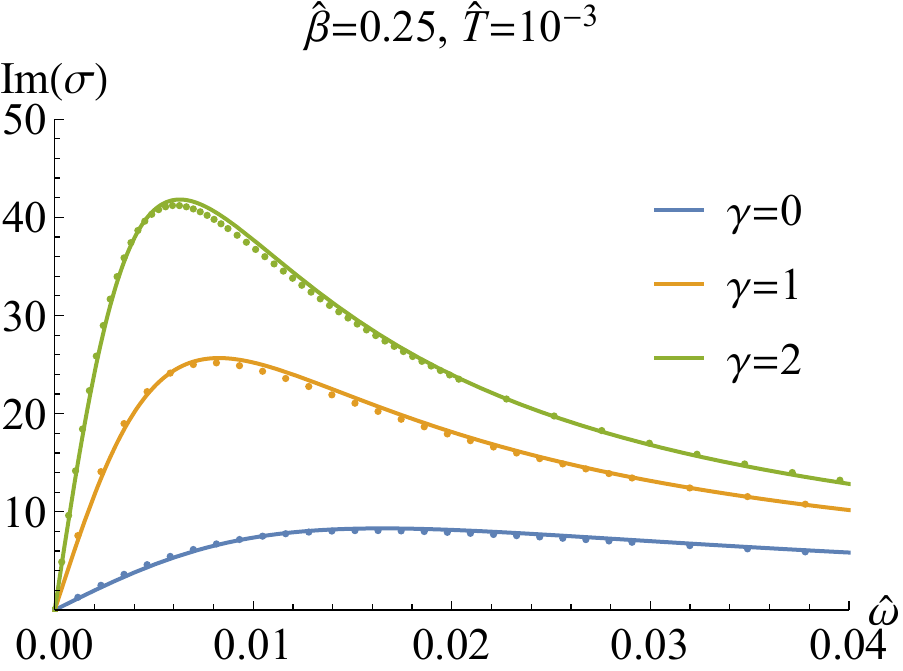}\ \hspace{0.8cm}\ \\
\caption{\label{sigma_alpha_0p25_f} The electric conductivity $\sigma(\hat{\omega})$ for different $\gamma$ and the fixed $\hat{\beta}=0.25$.
The dots are the numerical results while the solid lines are fitted by the standard Drude formula \eqref{Drude-s}.}}
\end{figure}
\begin{figure}
\center{
\includegraphics[scale=0.6]{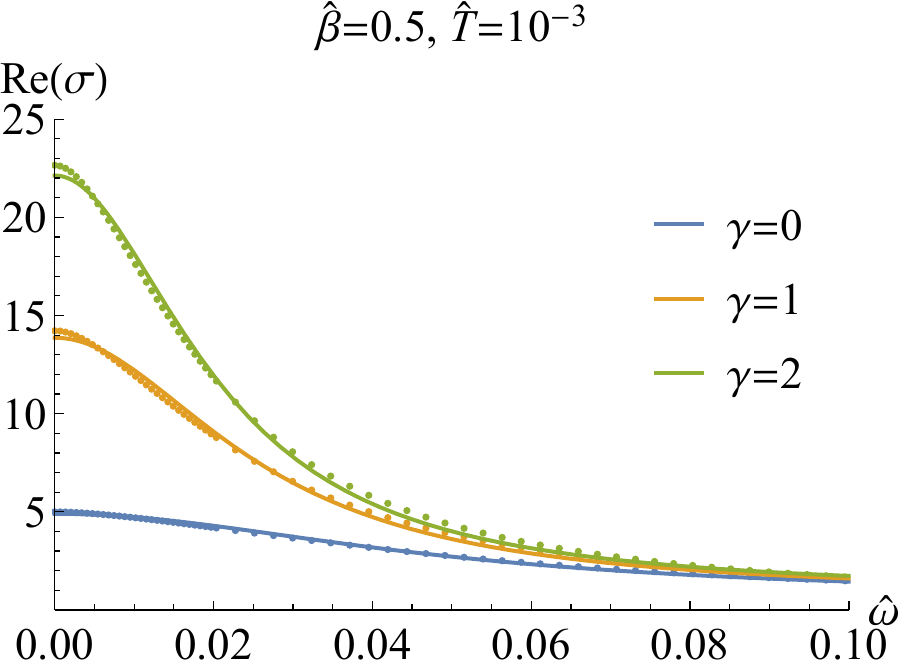}\ \hspace{0.8cm}
\includegraphics[scale=0.6]{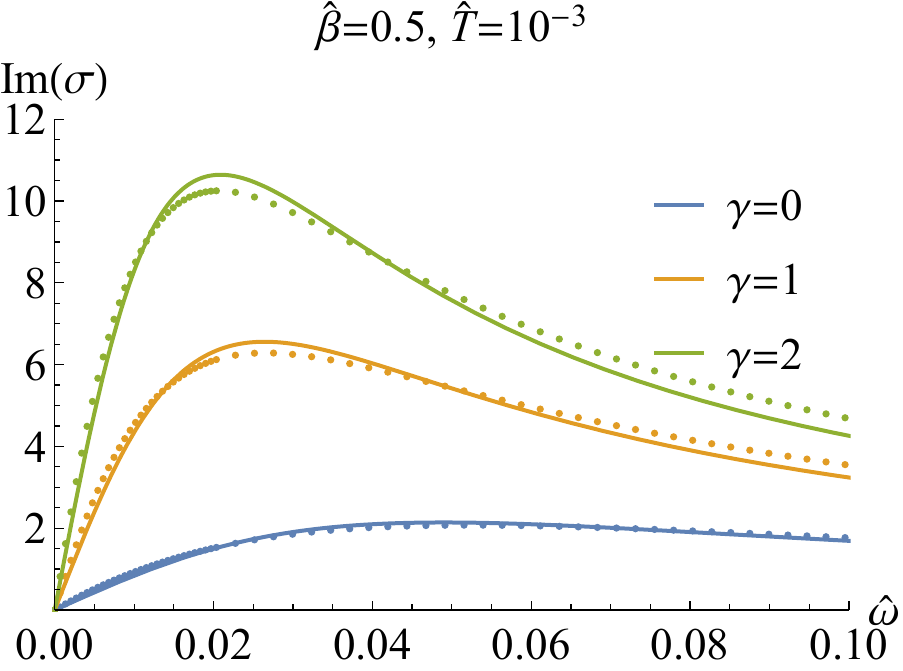}\ \hspace{0.8cm}\ \\
\includegraphics[scale=0.6]{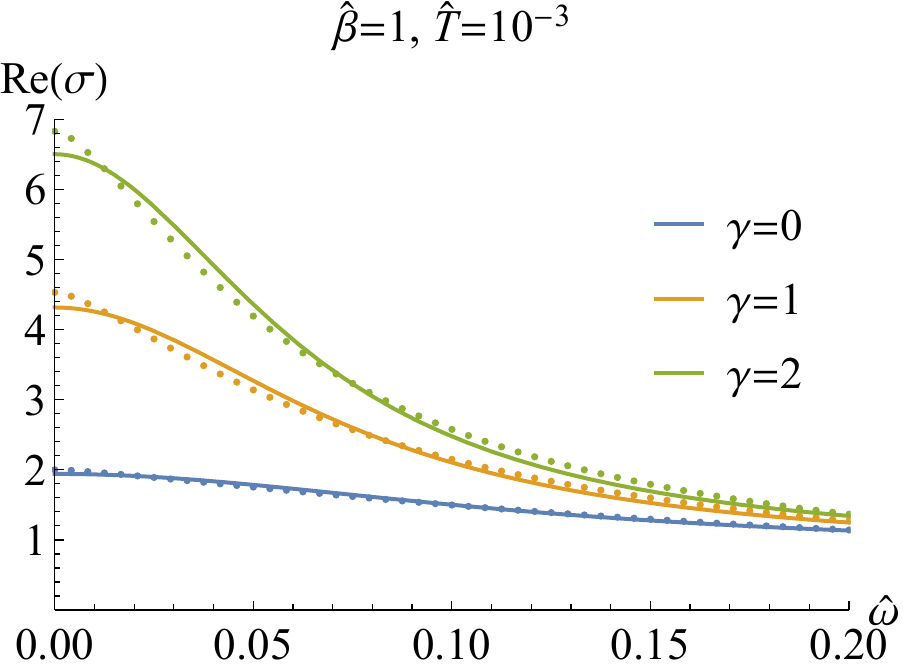}\ \hspace{0.8cm}
\includegraphics[scale=0.6]{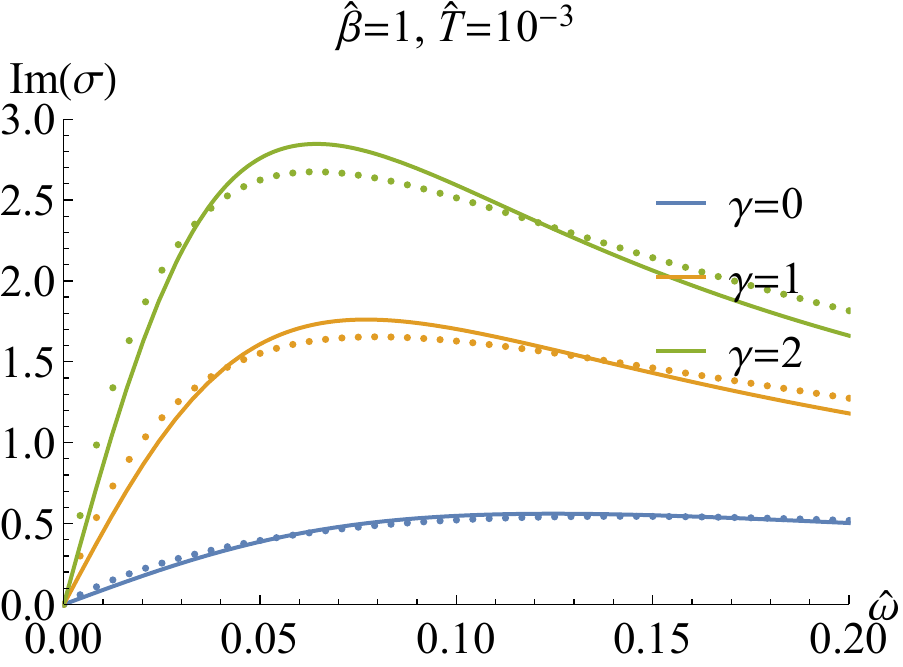}\ \hspace{0.8cm}\ \\
\caption{\label{sigma_alpha_0p5_1_f} The electric conductivity $\sigma(\hat{\omega})$ for different $\gamma$ and $\hat{\beta}$ (the plots above is for $\hat{\beta}=0.5$ and the ones below for $\hat{\beta}=1$).
The dots are the numerical results while the solid lines are fitted by the Modified Drude formula \eqref{Drude-m}.}}
\end{figure}

FIG.\ref{sigma_alpha_0p25_f} shows the electric conductivity $\sigma(\hat{\omega})$ for weak momentum dissipation ($\hat{\beta}=0.25$).
It can fitted very well by the standard Drude formula \eqref{Drude-s} and we conclude that when the momentum dissipation is weak,
the electric transport from BI-axions model is coherent for different BI coupling parameter $\gamma$.
Quantitatively, the momentum relaxation rate $\Gamma$ decreases with the increase of $\gamma$ (see TABLE \ref{table-fit-alpha-0p25}).
When $\hat{\beta}=0.5$, we need resort to the modified Drude formula \eqref{Drude-m} to fit the numerical data.
The results are shown in the above plots in FIG.\ref{sigma_alpha_0p5_1_f},
which is fitted very well. It indicates a crossover from coherent to incoherent phase begins to appear around $\hat{\beta}\simeq 0.5$.
Similarly with the weakly momentum dissipation case, the momentum relaxation rate $\Gamma$ also decreases with the increase of $\gamma$ in the crossover region (see TABLE \ref{table-fit-alpha-0p5}).
But we note that although the peak in $Re(\sigma)$ enhances, the $\sigma_Q$, the quantity characterizing the incoherent degree, increases with the increase of $\gamma$,
which indicates that  the BI coupling amplifies the incoherent behavior.
With the further increase of $\hat{\beta}$, the incoherent behavior becomes stronger (see the plots below in FIG.\ref{sigma_alpha_0p5_1_f} and TABLE \ref{table-fit-alpha-1}).

\begin{table}[h]
\center{
\begin{tabular}{|c|c|c|c|}\hline
$\gamma$&$0$&$1$&$2$
\\\hline
$\Gamma$&$0.0165$&$0.00827$&$0.00630$
\\\hline
 \end{tabular}
\caption{\label{table-fit-alpha-0p25} The momentum relaxation rate $\Gamma$ fitted by the standard Drude formula \eqref{Drude-s} for different $\gamma$ with fixed $\hat{\beta}=0.25$.}}
\end{table}
\begin{table}[h]
\center{
\begin{tabular}{|c|c|c|c|}\hline
$\gamma$&$0$&$1$&$2$
\\\hline
$\Gamma$&$0.0490$&$0.0264$&$0.0209$
\\\hline
$\sigma_Q$&$0.621$&$0.751$&$0.847$
\\\hline
 \end{tabular}
\caption{\label{table-fit-alpha-0p5} The momentum relaxation rate $\Gamma$ fitted by the modified Drude formula \eqref{Drude-m} for different $\gamma$ with fixed $\hat{\beta}=0.5$.}}
\end{table}
\begin{table}[h]
\center{
\begin{tabular}{|c|c|c|c|}\hline
$\gamma$&$0$&$1$&$2$
\\\hline
$\Gamma$&$0.125$&$0.0770$&$0.0644$
\\\hline
$\sigma_Q$&$0.819$&$0.792$&$0.806$
\\\hline
 \end{tabular}
\caption{\label{table-fit-alpha-1} The momentum relaxation rate $\Gamma$ fitted by the modified Drude formula \eqref{Drude-m} for different $\gamma$ with fixed $\hat{\beta}=1$.}}
\end{table}

\section{Non-linear  electric conductivity in probe limit}
As is discussed in \cite{Baggioli:2016oju}, the usual way to study non-linear conductivity is to show the non-linear current-voltage diagram, from which we may see the non-linear behavior of the electric conductivity. To get analytical control, we will work in the probe limit, i.e., we ignore the mixture  with the metric perturbation and keep the non-linear self -couplings of the Maxwell field as done in the references \cite{Sonner:2012if,Horowitz:2013mia}.

We obtained in previous sections that $\sigma_{DC}$ is constant for fixed bulk parameters, that is to say, the DC conductivity is linear. Here we shall discuss the non-linear DC case  via the steps in \cite{Baggioli:2016oju}. Consider the gauge field as $A=A_t(u)dt+(E_xt +a_x(u))dx$, the Maxwell equations with the same form as \eqref{metric} give us
\begin{equation}
\dot{\mathcal{L}_{BI}} A_t'=-\rho,~~~\dot{\mathcal{L}_{BI}}f(u) a_x'=-J_x
\end{equation}
where $\dot{\mathcal{L}_{BI}}$ denotes the derivative to the field strength  $I=-F^2/2$ and prime denotes the derivative to the radius $u$ .  $\mathcal{L}_{BI}$ has been defined in \eqref{LM} while the integration constants $\rho$ and $J_x$ are interpreted as charge density and charge current. Further requiring the field strength
\begin{equation}
I=u^4(A_t^{'2}+\frac{E_x^2}{f(u)}-f(u)a_x^{'2})
\end{equation}
is regular at the horizon, we  obtain the genera l current- voltage relation
\begin{equation}\label{C-V}
J_x= \dot{\mathcal{L}_{BI}}\mid_{u_h} E_x.
\end{equation}

\begin{figure}
\center{
\includegraphics[scale=0.4]{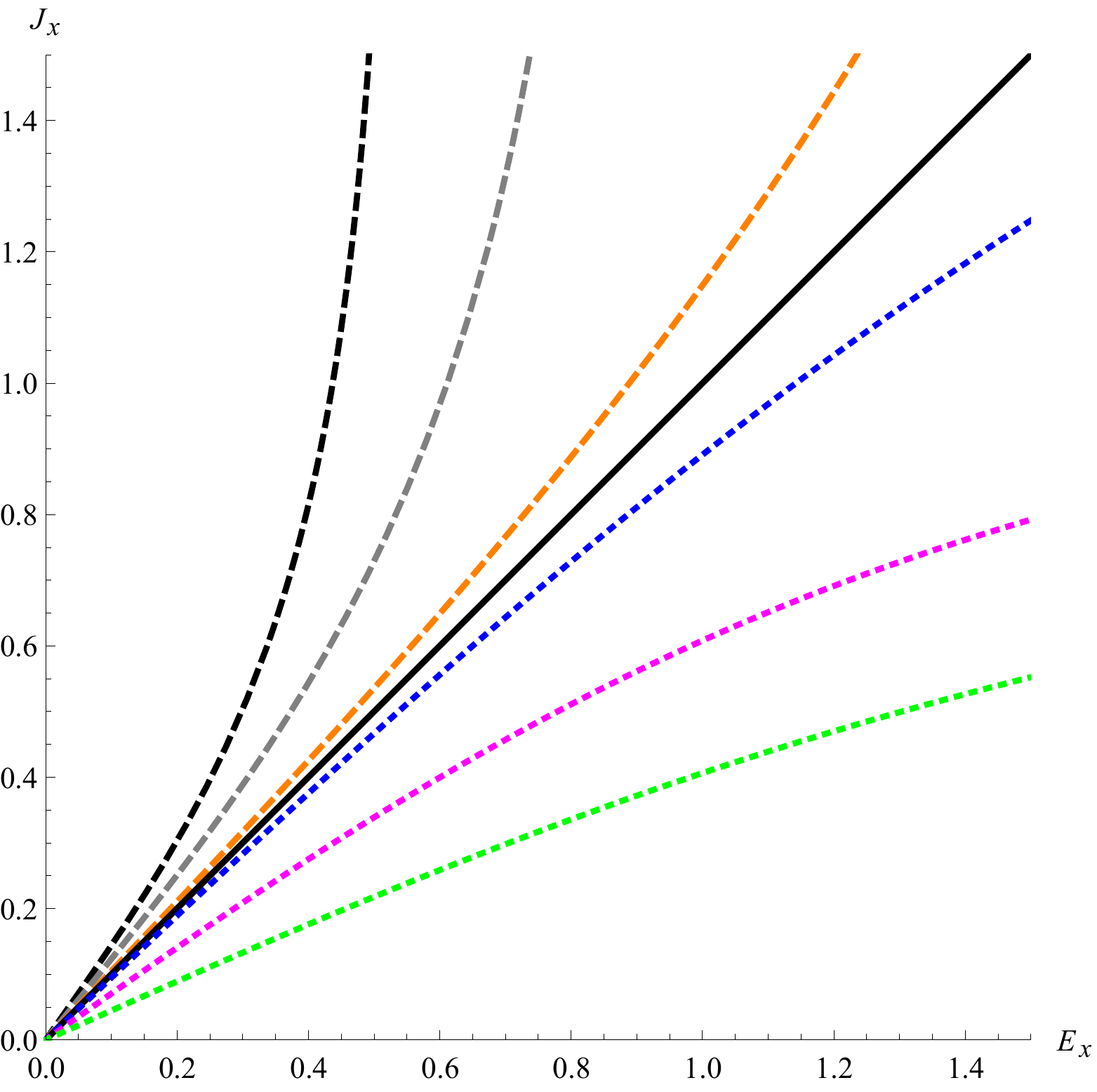}
\caption{\label{nonlinearConduc}The current-voltage behavior for BI model. We set $\mu=u_h=1$, and the parameter  $\gamma$ from top to down is $1,0.5,0.1,0,-0.1,-0.5,-0.8$.}}
\end{figure}
The current- voltage behavior from \eqref{C-V} is shown in FIG. \ref{nonlinearConduc}.  $\gamma=0$ corresponds to the standard Maxwell theory, so that the current- voltage relation  is linear and means $J_x=E_x$ as we expect. When $\gamma\neq 0$, the non-linear
behavior is observed.  For  $\gamma> 0$, the curve is above the linear case and $dJ_x/dE_x$ is stronger than $1$ as that happened in DBI model\cite{Baggioli:2016oju}, however, it is enhanced as the applied field $E_x$ increases which is different from that in DBI model. For $\gamma<0$, the non-linear $dJ_x/dE_x$  is always lower than $1$.  As the applied field increases, it approaches to be vanished, but this will not happen because backreacion should be involved  when $E_x$ is very large . The unstable case $dJ_x/dE_x<0$ shown in iDBI model  is not observed in our model. We notice that when we continue to lower $\gamma$, the current become complex, this deserves further study.

\section{Conclusions}

In this paper, we introduced the Maxwell field with Born-Infeld correction into the Einstein-axions theory and constructed a new charged BI-AdS black hole. Then we analytically calculated  various
DC transport coefficients of the dual boundary theory. We found that the DC electric conductivity
depends on the temperature of the boundary theory, which is a novel property comparing to that in RN-AdS black hole. At zero temperature,
The DC electric conductivity are positive while the thermal conductivity vanishes. This means that the dual sector is an electrical metal but thermal insulator. With the increase of Born-Infeld parameter, the electric conductivity, electric-thermo conductivity and thermal conductivity at zero increases at finite fixed temperature.

We also studied the AC electric conductivity of the theory. When the momentum
dissipation is weak, the low frequency AC conductivity behaves as the standard Drude formula and the electric transport is coherent for various correction parameter. When the momentum dissipation is stronger, the modified Drude formula is applied and a crossover from coherent to incoherent phase was observed.
Also, we found that the Born-Infeld correction makes the incoherent behavior more explicit. We notice that here we only numerically compute the AC electric conductivity dual to its simply. It would be very interesting to further study the AC thermal and electric-thermo conductivity which are related with boundary data of both Maxwell perturbation and  gravitational perturbation\cite{Amoretti:2014zha}. We hope to show the results elsewhere soon.

Finally, we analyze the non-linear current-voltage behavior with BI correction in probe limit. The curve with $\gamma> 0$ is above the linear case and $dJ_x/dE_x$ is always  bigger than $1$. Different from that happened in DBI model \cite{Baggioli:2016oju},  the slope is enhanced as $E_x$ increases. For $\gamma<0$, the non-linear $dJ_x/dE_x$  is always lower than $1$ and it tends to be zero as $E_x$ go to infinity in which case the backreaction should be considered.  For more negative $\gamma$, the current become complex and further study is called for.

There are many interesting questions deserving further exploration.
First of all, we can study the holographic anomalous transport from BI electrodynamics as \cite{Chu:2018ksb,Chu:2018ntx}.
In \cite{Li:2017nxh,Mokhtari:2017vyz}, they study the thermal transport and quantum chaos in
the EMDA theory with a small Weyl coupling term. In particular, in \cite{Li:2017nxh},
they find that the Weyl coupling correct the thermal diffusion constant and butterfly velocity in different ways, hence resulting in a
modified relation between the two at IR fixed points.
It is interesting to further explore this relation in present of BI correction.
We shall come back these topics in near future.

\begin{acknowledgments}
We are very grateful to Peng Liu for many useful discussions and comments on the manuscript.
This work is supported by the Natural Science
Foundation of China under Grants No. 11705161, No. 11775036 and  No. 11747038.
X. M. Kuang is also supported by Natural Science Foundation of Jiangsu Province under Grant No.BK20170481.
J. P. Wu is also supported by the Natural Science Foundation of Liaoning Province under Grant No.201602013.
\end{acknowledgments}

\end{document}